\documentclass[hyper]{PoS}

\usepackage{graphicx}
\usepackage{arydshln}
\DeclareGraphicsExtensions{.pdf,.png,.jpg}
\usepackage{amsmath}
\newcommand\slurp[1]{#1}
\newcommand\sslurp[2]{#1{#2}}
{\catcode`/=\active \expandafter}%
\slurp{\newcommand/}{/\penalty1000\hskip0pt\relax}
{\catcode`:=\active \expandafter}%
\slurp{\newcommand:}{:\penalty1000\hskip0pt\relax}

\newcommand\addspace{\ifcat\nextchar a\spacefactor999. \else.\fi}
{\catcode`\.=\active \expandafter}%
\slurp{\newcommand.}{\futurelet\nextchar\addspace}

\ifx\href\undefined\def\href#1{}\fi
\ifx\texorpdfstring\undefined\def\texorpdfstring#1#2{#1}\fi

\def\parsemultipledoi#1, {\dodoi{#1}\futurelet\nexttok\checktok}
\def\checktok{\ifx\nexttok\relax\else, \expandafter\parsemultipledoi\fi}
\newcommand\doi[1]{\parsemultipledoi#1, \relax}
\newcommand\myslash{/} \newcommand\mycolon{:}
\newcommand\dodoi{{\catcode`/=\active \catcode`:=\active \expandafter}\sslurp\realdoi}
{\catcode`/=\active \catcode`:=\active \expandafter}%
\slurp{\newcommand\realdoi[1]{{\let/=\myslash \let:=\mycolon
                               \edef\raw{{http://dx.doi.org/#1}}\expandafter}%
                               \expandafter\href\raw{doi:#1}}}
\newcommand\eprint[2]{{\escapechar-1%
                       \edef\a{\expandafter\string\csname arXiv\endcsname}%
                       \edef\b{\expandafter\string\csname #1\endcsname}%
                       \edef\c{\expandafter\string\csname #2\endcsname}%
                       \edef\d{\noexpand\href{http://arXiv.org/abs/\c}}%
                       \ifx\a\b\expandafter\d\fi{\tt #1:#2}}}

\newcommand\slashnext[1]{{\setbox0\hbox{\(#1\)}\setbox1\hbox to \wd0{\hss\it/\hss}\rlap{\box1}{\box0}}}
\newcommand{\Tslash}{\slashnext T}
\newcommand{\Aslash}{\slashnext A}
\newcommand{\Dslash}{\slashnext D}
\newcommand{\qslash}{\slashnext q}
\newcommand{\MSbar}{$\overline{\rm MS}$}

\title{Neutron Electric Dipole Moments from Beyond the Standard Model Physics}

\ShortTitle{Neutron Electric Dipole Moment from Beyond the Standard Model Physics}

\author{\speaker{Tanmoy Bhattacharya}%
        \\
        Los Alamos National Laboratory\\
        E-mail: \email{tanmoy@lanl.gov}}

\author{Vincenzo Cirigliano\\
        Los Alamos National Laboratory\\
        E-mail: \email{cirigliano@lanl.gov}}

\author{Rajan Gupta\\
        Los Alamos National Laboratory\\
        E-mail: \email{rajan@lanl.gov}}

\abstract{Neutron Electric Dipole Moment  (nEDM), a generic feature of CP-violation, is 
predicted to be very small in the Standard Model, but can be much larger in most 
extensions of the model. In this talk, I will discuss the classification of the CP 
violating operators up to dimension 6 that can give rise to nEDM, and then describe
 the   mixing and   renormalization structure of the operators of dimension 5 and lower 
 in both dimensional and cutoff   regularizations in general terms. Finally I will describe how to connect the 
dimension 5 operators, in particular, the Chromoelectric Dipole Moment of the 
quarks, between MSbar scheme and a Regularization Independent prescription in 
the chiral limit.}

\FullConference{31st International Symposium on Lattice Field Theory LATTICE 2013\\
                 July 29--August 3, 2013\\
                 Mainz, Germany}

\begin{document}

\section{Introduction}

Electric dipole moments (EDMs) of non-degenerate elementary particles violate both parity (P) and time reversal (T) symmetries. A violation of time-reversal symmetry is necessary for generating the observed baryon-asymmetry in the standard cosmological scenario~\cite{CPnecessary}.  The standard model of particle physics violates this symmetry through the imaginary part of the Cabibbo-Kobayashi-Maskawa (CKM) mixing matrix~\cite{CKM} in the weak interaction sector, but this is insufficient to generate enough baryon asymmetry~\cite{toolittle}.  Most extensions of the standard model allow P and T violations and may give rise to EDMs for elementary particles.  Searching for such EDMs may, therefore, be a fruitful way for understanding this sector of particle physics~\cite{EDMgood}.  In this talk, I will focus on the neutron EDM (nEDM) where rapid experimental progress is expected~\cite{experiments}.

In principle, the standard model contains two possible sources of T violation.  The contribution of the complex phase in the CKM matrix is always suppressed by terms proportional to the quark mixing and the quark masses; the nEDM due to this has been estimated to be no more than \(10^{-32} \mbox{e-cm}\)~\cite{Dar}.  The other source of CP violation is from the QCD instanton density whose contribution is suppressed only by quark masses, so the unobserved nEDM constrains the corresponding coefficient, \(\Theta\), in the Lagrangian  to be unnaturally small, \(\Theta \lesssim 10^{-10}\)~\cite{Crewther}.  Because of this, it is most often assumed that this term relaxes to zero dynamically~\cite{PQ}.

\section{Effective Field Theory}

To study the physics of T violation systematically, we follow the effective field theory (EFT) approach. For simplicity we will be writing our formulae for a \(N_f=2\) flavor theory, but its generalizations to an arbitrary \(N_f > 1\) are simple.

\subsection{Dimensions 3 and 4}
The renormalizable terms, mass-terms of dimensions 3 and the instanton contribution of dimension 4, that violate T can be written as follows:
\begin{equation}
S^{(4)}_{\Tslash} \subset -  \int d^4x  \left[   \overline\psi (m + i m_5\gamma_5) \psi 
                   + \overline\psi (m^I + i m_5^I \gamma_5) \tau_3 \psi + \Theta \frac{g^2}{16\pi^2}\epsilon_{\alpha\beta\mu\nu}G^{\alpha\beta\,A}G^{\mu\nu\,A}\, \right]~, 
\end{equation}
where \(G\) represents the color octet gluonic field strength, \(\psi\) the fermion field that belongs to the fundamental representation of both color SU(3) and flavor SU(2) groups, and \(\tau_3\) is the diagonal generator of the flavor group. In the standard model, the U(1) instantons do not contribute to any finite action configurations, and, due to the anomaly, the effect of the weak instantons can be absorbed by a phase choice since only the left-chiral quarks are charged under weak interactions whereas the rest of the theory is invariant under equal and opposite rotations of the left and right chiral quarks.

\subsection{Chiral rotations}
T-violating Lagrangian contains four mass terms in addition to the coefficient, \(\Theta\). Under chiral rotations of the quark fields
\begin{subequations}
\begin{eqnarray}
   \psi&\to&e^{i\gamma_5\alpha} \psi \\
   \psi&\to&e^{i\gamma_5\beta\tau_3} \psi\,.
\end{eqnarray}
\end{subequations}
the mass terms and the coefficient \(\Theta\) transform as
\begin{equation}
\left(\begin{array}{l} m\\m_5\\m^I\\m_5^I\\\Theta \end{array}\right) =
   \left(\begin{array}{rrrrc}
         \cos\alpha\cos\beta & -\sin\alpha\cos\beta & -\sin\alpha\sin\beta & -\cos\alpha\sin\beta & 0 \\
         \sin\alpha\cos\beta & \cos\alpha\cos\beta & \cos\alpha\sin\beta & -\sin\alpha\sin\beta & 0 \\
         -\sin\alpha\sin\beta & -\cos\alpha\sin\beta & \cos\alpha\cos\beta & -\sin\alpha\cos\beta & 0\\
         \cos\alpha\sin\beta & -\sin\alpha\sin\beta & \sin\alpha\cos\beta & \cos\alpha\cos\beta & 0\\
         \multicolumn1c0&\multicolumn1c0&\multicolumn1c0&\multicolumn1c0&\multicolumn1c1\end{array}\right)
   \left(\begin{array}{l}m\\m_5\\m^I\\m_5^I\\\Theta+\alpha \end{array}\right)\,,
\end{equation}        
and the rest of the Lagrangian is invariant. The manifold of theories that are inequivalent under these reparametrizations is the quotient space \({\mathbb{C}}^{N_f}\times{\mathbb{S}}^1 / U(N_f)\) and has a conical singularity whenever more than one mass is zero\rlap{.}\footnote{The standard representation that makes the masses real has an additional coordinate singularity when any mass is zero.}

\subsection{Dimension 5}
\begin{figure}[t]
\begin{center}
\includegraphics[width=\textwidth]{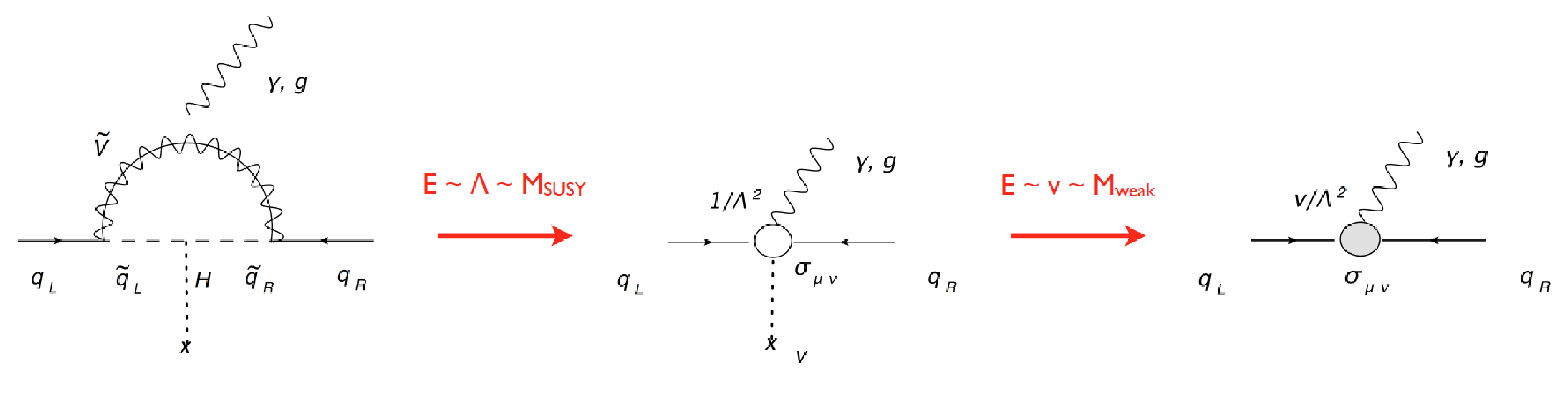}
\end{center}
\caption{One loop diagrams in beyond the Standard Model physics give rise to quark EDM and chromo-EDM.\label{EDM}}
\end{figure}

At dimension five, there are two kinds of CP violating operators: the quark electric and chromo-electric dipole moments:
\begin{eqnarray}
S^{(5)}_{\Tslash} &=& i e (d^\gamma \overline\psi F^{\mu\nu}\sigma_{\mu\nu} \gamma_5 \psi + 
                           d_I^\gamma \overline\psi F^{\mu\nu} \sigma_{\mu\nu} \gamma_5 \tau_3 \psi) \nonumber\\
                  &&{} + i g_s (d^G \overline\psi G^{\mu\nu}\sigma_{\mu\nu} \gamma_5 \psi + 
                           d_I^G \overline\psi G^{\mu\nu} \sigma_{\mu\nu} \gamma_5 \tau_3 \psi)\,,\label{eq5}
\end{eqnarray}
where \(e\) and \(g_s\) are the electromagnetic and strong coupling constants, \(d^{\gamma,G}\) are the iso-scalar quark EDM and quark chromo-EDM and \(d_I^{\gamma,G}\) are the corresponding iso-vector quantities.  These arise from dimension six operators involving the Higgs' field above the the weak SU(2)$_{\rm W}$ breaking scale, and hence they are suppressed by \(v/\Lambda_{\rm BSM}^2\), where \(v\) is the SU(2)$_{\rm W}$ breaking vacuum expectation value, and \(\Lambda_{\rm BSM}\) is the scale of new physics responsible for these operators.  Note that since they are vectors under the chiral symmetry, in many fundamental theories they appear proportional to the quark masses that have the same symmety structure.  They arise at 3-loops in the standard model from the CKM phase, and their contribution is extremely small, about \(10^{-34} \mbox{e-cm}\)~\cite{Dar}. In a generic beyond-the-standard-model (BSM) theory, they arise at one loop (see Figure~\ref{EDM}), and can give an EDM large as the current experimental limit of approximately \(10^{-26}\mbox{e-cm}\)~\cite{Baker}. Note that unless there is an alignment of the dimension 5 and dimension 4 terms, a chiral rotation to make the quark masses real will, in general, produce these T-violating EDMs from any T-conserving anomalous magnetic dipole moments in the BSM theory.  

\subsection{Dimension 6}
Since the dimension five operators in Eq.~(\ref{eq5}) have a \(v/\Lambda^2\) suppression, one should analyze the effect of T-violation from dimension six operators as well.  At low energies, these appear as the chromoelectric dipole moment of the gluon represented by the Weinberg operator~\cite{Weinberg:1989dx}
\begin{subequations}
\begin{equation}
 \frac13 f^{ABC} \epsilon_{\nu\alpha\beta\phi} G^{\mu\nu\,A}G^{\alpha\beta\,B}G^{\hphantom{\mu}\phi\,C}_\mu\,.
\end{equation}
In addition, there are four-fermion operators of various kinds
\begin{eqnarray}
&&\frac i4 \epsilon^{jk}\overline\psi^j u \overline\psi^k d\\
&&\frac i4 \epsilon^{jk}\overline\psi^j T^a u \overline\psi^k T^a d\\
&&\frac i2 \psi^j\gamma^\mu\psi^k \overline u \gamma_\mu d\,,
\end{eqnarray}
\end{subequations}
where \(\psi^{j,k}\) represent various flavors of left chiral fields, and \(u,d\) similarly represent right chiral fields, and \(T^a\) is the color generators in the fundamental representations.~\cite{dim6ops1,dim6ops2,dim6ops3,Hisano:2012cc}.

\section{Operator Mixing}

\begin{figure}[t]
\begin{center}
\begin{tabular}{cc}
\raise50pt\hbox{
\includegraphics[clip=true,trim=0pt 0pt 0pt 350pt,width=0.3\hsize]
                {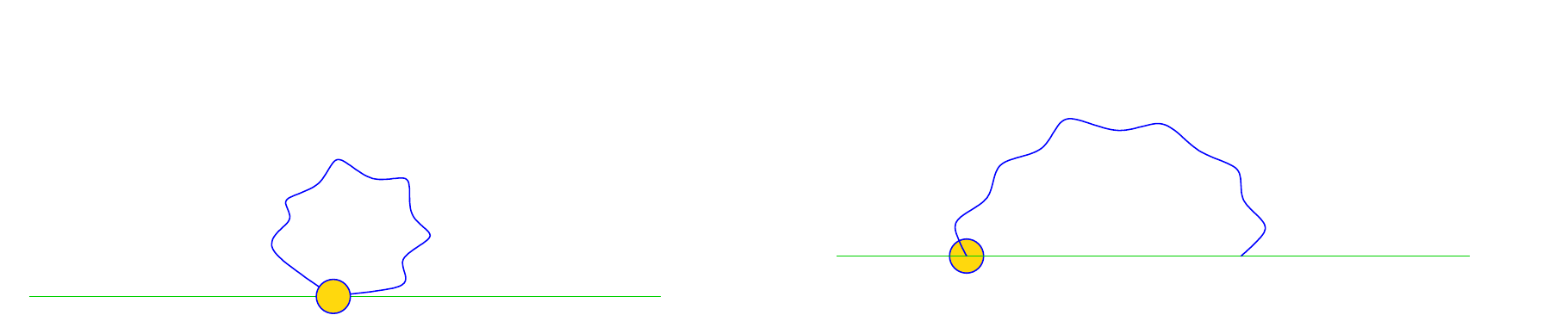}} &
\includegraphics[width=0.35\hsize]{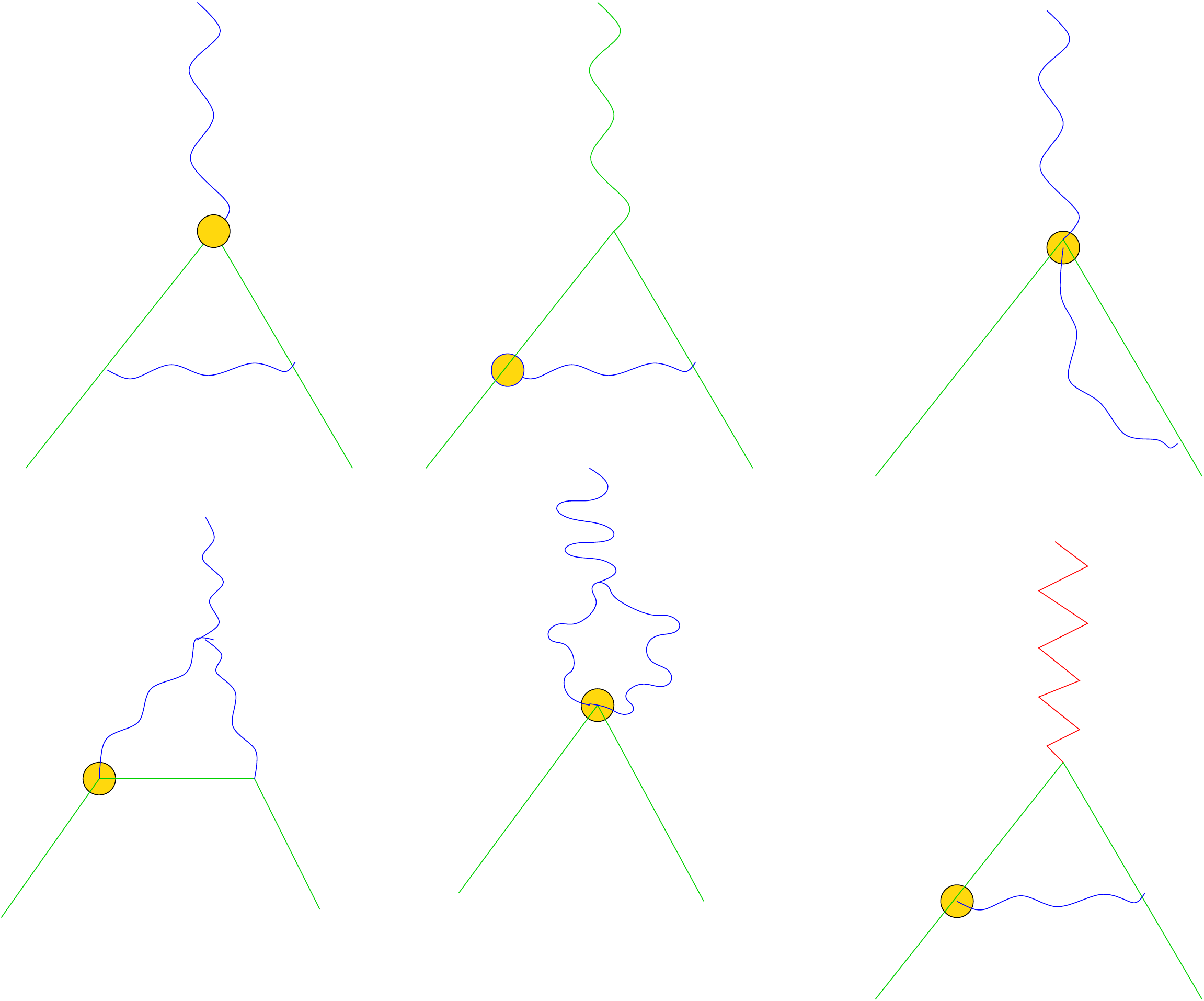}\\
(a)&(b)
\end{tabular}
\end{center}
\caption{The superficially divergent Feynman diagrams with (a) two and (b) three external legs.  The wavy lines represent gluons, the angular lines the photon and the straight lines the quark.  A circled vertex is CP-violating. \label{irreduc}}
\end{figure}

The 1-particle irreducible diagrams needed to renormalize these operators are shown in Figure~\ref{irreduc}.  In the \MSbar\ scheme and at zeroth order in the elecromagnetic coupling, the operators at dimensions 3, 4, and 5 are all multiplicatively renormalized in the chiral limit except for the quark chromo-EDM\rlap{.}\footnote{In the strict chiral limit, each of the dimension five T-violating isovector operators can be rotated away to a T-conserving one by a phase choice of the fermion fields. One can do the same for the T-violating isoscalar operators at the expense of changing the \(\Theta\)-term.}{\spacefactor\sfcode`\.}  This is because in massless \MSbar\ scheme, operators of different dimensions do not mix.  The two operators at dimension three are different under the flavor symmetry group, and the quark EDM operators involving a photon cannot mix with photon-free operators without a photon-loop.  The only mixing, then, that one needs to consider is the quark chromo-EDM operator can induce a quark EDM at one loop. 

In a scheme that allows mixing between operators of different dimensions, the chromo-EDM operator can mix with the mass operators. As already noted, these operators can be rotated away by a chiral rotation at the expense of possibly changing the \(\Theta\)-term.  Such a chiral rotation, however, also mixes the chromomagnetic and chromoelectric dipole moment operators, which differ by a \(\gamma_5\). 

\subsection{RI-sMOM scheme}
\begin{figure}[t]
\begin{center}
\begin{tabular}{cc}
\includegraphics[width=0.3\hsize]{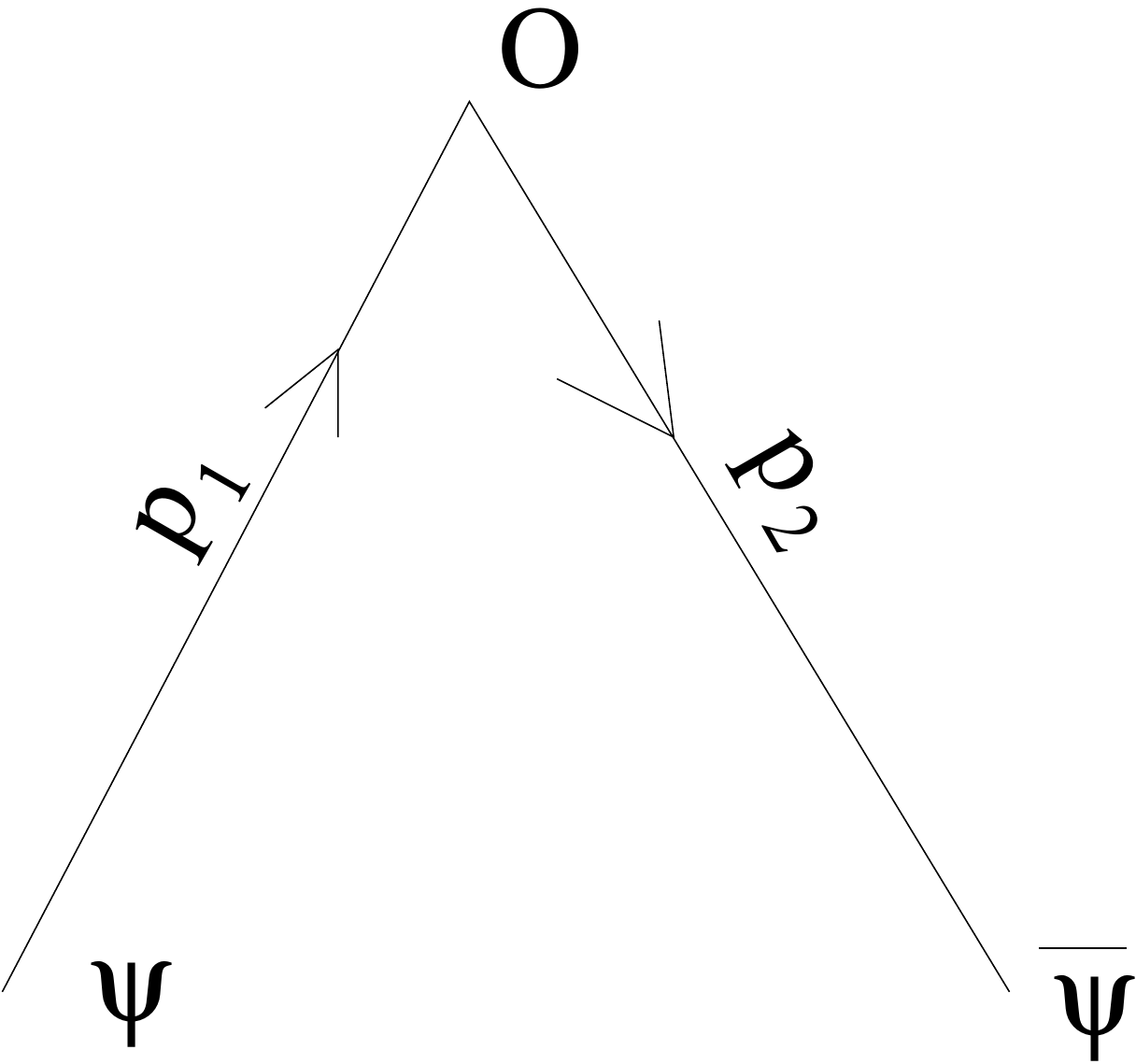}&
\includegraphics[width=0.3\hsize]{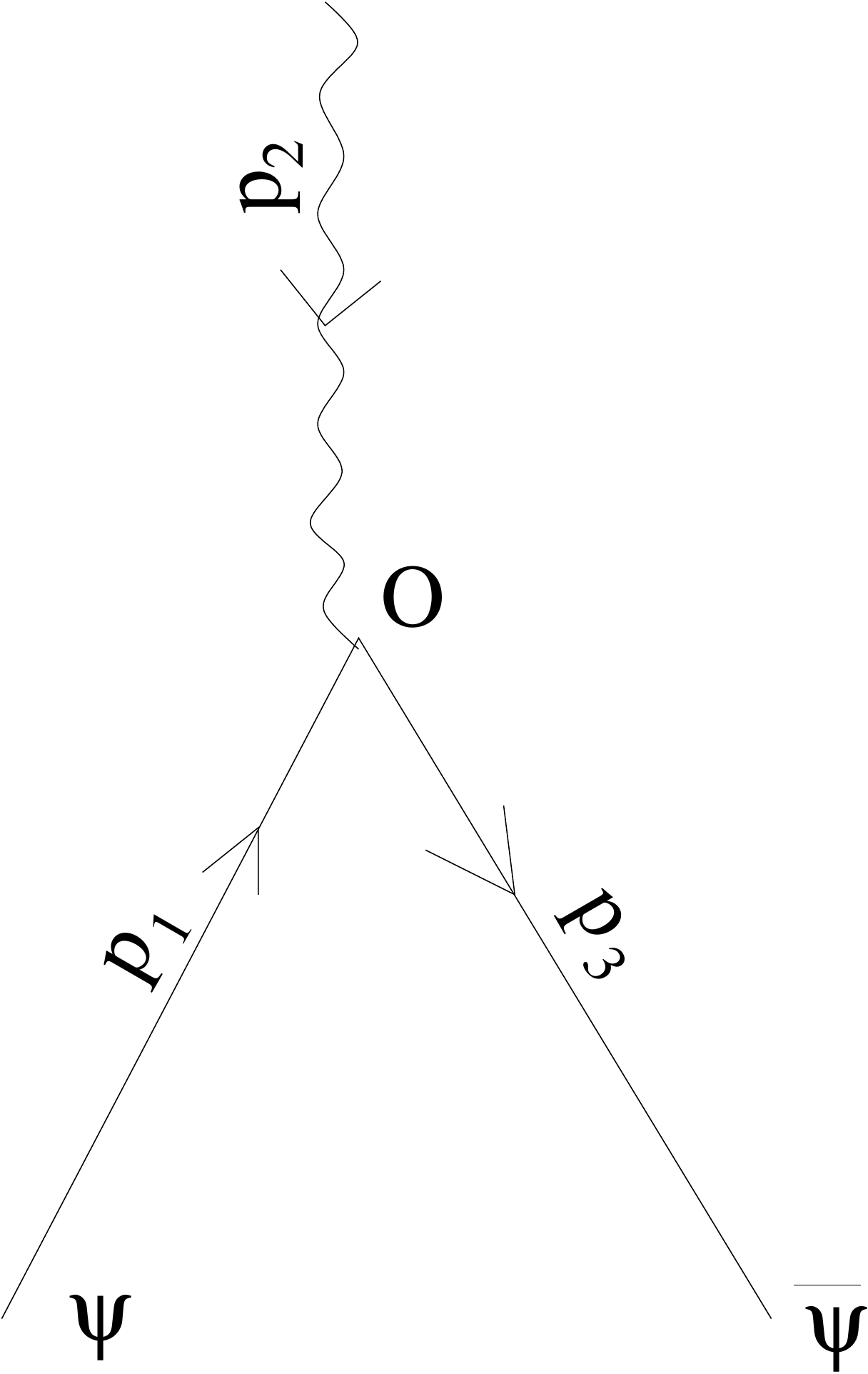}\\
(a)&(b)
\end{tabular}
\end{center}
\caption{The momentum flow in matrix elements with (a) two and (b) three external legs in the RI-sMOM scheme.\label{RIsMOM}}
\end{figure}
To match between the purely perturbative \MSbar\ lattice scheme and the lattice regularized theory, it is convenient to calculate a perturbatively calculable non-exceptional matrix element in the deep Euclidean region.  This is usually done by constructing the RI-sMOM scheme that considers the three point function of the operator inserted in external quark or gluon states at the symmetric momentum point (see Figure~\ref{RIsMOM})
\begin{equation}
p_1^2 = p_2^2 = (p_2-p_1)^2 \equiv \Lambda^2\,,
\end{equation}
where \(\Lambda\) plays the role of the renormalization scale.  For operators with three external legs, we can continue this approach by choosing 
\begin{subequations}
\begin{eqnarray} 
&& p_1^2 = p_2^2 = p_3^2 = (p_3-p_1-p_2)^2 \equiv \Lambda^2\\
&& (p_1+p_2)^2 = (p_3-p_1)^2 = (p_3-p_2)^2 \equiv \frac43\Lambda^2\,.
\end{eqnarray}
\end{subequations}

\subsection{Off-shell BRST symmetry}
The problem with matching such truncated Green's function is that they become perturbative only in the deep Euclidean region, and, so, the calculation needs to be done in momentum space in a fixed gauge.  Renormalization in a off-shell gauge-fixed theory allows mixing with gauge-variant and equation-of-motion operators.  These extra operators do not contribute to physical matrix elements~\cite{DeansDixon}, but they do affect the matching calculations.  The allowed operators are, however, restricted by the invariance of the gauge fixed theory under the Becchi-Rouet-Stora-Tyutin (BRST) transformations~\cite{BRST}.

The application of off-shell BRST symmetries~\cite{DeansDixon} to the Landau gauge fixed theory imply that there are no extra dimension 3 operators.  New operators do indeed appear at dimensions 4 and 5. The T-violating operators up to dimension 5 can then be rearranged into a convenient basis given in Table~\ref{tree}.
\begin{table}[t]
\begin{center}
\begin{tabular}{|l|l|l|}
\hline
\multicolumn{1}{|c|}{Operator}&\multicolumn{1}{c|}{\(\langle q|O|q\rangle\)}&\multicolumn{1}{c|}{1PI \(\langle qg | O | q\rangle\)}\\[1\jot]
\hline
\strut\vphantom{\({}^2\)}\(\overline\psi i\gamma_5 \psi\)&\(i\gamma_5\)&0\\[1\jot]
\hline
\strut\vphantom{\({}^2\)}\(G_{\mu\nu}\tilde G^{\mu\nu}\)&0&0\\[1\jot]
\hdashline
\strut\vphantom{\({}^2\)}\(\overline\psi_E i\gamma_5\psi + \overline\psi i\gamma_5\psi_E\)&\(i(\qslash\gamma_5 - 2 m \gamma_5)\)&0\\[1\jot]
\hline
\strut\vphantom{\({}^2\)}\(\overline\psi\tilde\sigma^{\mu\nu}F_{\mu\nu}\psi\)&0&0\\[1\jot]
\strut\vphantom{\({}^2\)}\(\overline\psi\tilde\sigma^{\mu\nu}G_{\mu\nu}\psi\)&0&\(-2i\tilde\sigma^{\rho\nu}k_\nu\)\\[1\jot]
\hdashline
\strut\vphantom{\({}^2\)}\(i\overline\psi_E\gamma_5\psi_E\)&\(-i[m\qslash\gamma_5 +(p'\cdot p -m^2)\gamma_5]\)&\(-g[i(p+p')^\rho\gamma_5]+\tilde\sigma^{\rho\nu}(k-q)_\nu]\)\\
&\multicolumn{1}{r|}{\(\qquad{}+p_\mu\tilde\sigma^{\mu\nu}p'_\nu\)}&\\[1\jot]
\strut\vphantom{\({}^2\)}\(i\partial_\mu[\overline\psi_E\gamma^\mu\gamma_5\psi+\overline\psi\gamma^\mu\gamma_5\psi_E]\)&\(-2m\qslash\gamma_5+q^2\gamma_5-2ip_\mu\tilde\sigma^{\mu\nu}p'_\nu\)&\(-2ig\tilde\sigma^{\rho\nu}q_\nu\)\\[1\jot]
\strut\vphantom{\({}^2\)}\(\overline\psi\Aslash\gamma_5\psi_E+\overline\psi_E\Aslash\gamma_5\psi\)&0&\(q^\rho\gamma_5 + i\tilde\sigma^{\rho\nu}(p+p')_\nu-2m\gamma^\rho\gamma_5\)\\[1\jot]
\strut\vphantom{\({}^2\)}\(i\partial_\mu\overline\psi\tilde\sigma^{\mu\nu}i\overleftrightarrow D_\nu\psi\)&\(-2p_\mu\tilde\sigma^{\mu\nu}p'_\nu\)&\(-2g\tilde\sigma^{\rho\nu}q_\nu\)\\[1\jot]
\strut\vphantom{\({}^2\)}\((\partial^\mu \overline c)D_\mu c\)&0&0\\[1\jot]
\strut\vphantom{\({}^2\)}\(\partial_\mu [(\partial^\mu\overline c) c]\)&0&0\\[1\jot]
\hline
\end{tabular}
\end{center}
\caption{A convenient basis for T-violating operators up to dimension five and their truncated tree-level matrix elements.  Here, we use \(\psi_E \equiv (i\Dslash - m) \psi\), \(D\) for the covariant derivative, \(c,\overline c\) for the Fadeev-Popov (FP) ghosts, and we have suppressed the flavor structures. In each block, the operators below the dashed lines do not contribute to physical matrix elements at zero momentum insertion.\label{tree}}
\end{table}
The additional operators fall into three classes: operators including FP ghosts, those that vanish by equations of motion, and those that are total derivatives. None of these operators contribute to the nEDM, but the latter two do contribute to the RI-sMOM correlators.

Having identified the relevant operators, we are calculating the renormalization constants in both $\overline{\rm MS}$ and RI-sMOM schemes. 
We have computed the two- and three-point functions at tree level and are currently completing the calculation of the loop diagrams depicted in Fig.~\ref{irreduc}, 
using the 't Hooft-Veltman scheme for  $\gamma_5$  in dimensional regularization. The full results will be reported in a separate publication.

\section{Acknowledgements}

We acknowledge DOE grant DE-KA-2401012 and  funding from the LDRD program at LANL.  

\bibliographystyle{doiplain}
   \let\oldnewblock=\newblock
    \newcommand\dispatcholdnewblock[1]{\oldnewblock{#1}}
    \renewcommand\newblock{\spaceskip=0.3emplus0.3emminus0.2em\relax
                           \xspaceskip=0.3emplus0.6emminus0.1em\relax
                           \hskip0ptplus0.5emminus0.2em\relax
                           {\catcode`\.=\active
                           \expandafter}\dispatcholdnewblock}
\bibliography{main}

\end{document}